\documentclass[a4paper,11pt]{article}

\usepackage{jheppub} 

\usepackage[T1]{fontenc} 

\usepackage[utf8]{inputenc}
\usepackage{url}
\usepackage{graphicx}
\usepackage{axodraw2}

\title{
\normalfont
\vskip-2cm{\baselineskip14pt
  \begin{flushleft}
      \normalsize HU-EP-13/04\\
      \normalsize HU-Mathematik:05-2013 
  \end{flushleft}}
  \vskip1.5cm
  \boldmath FIRE4, LiteRed and accompanying tools to solve integration by parts relations}

\author[a]{Alexander V. Smirnov}
\author[b,c]{Vladimir A. Smirnov}

\affiliation[a]{Scientific Research Computing Center,\\Moscow State University,\\119992 Moscow, Russia}
\affiliation[b]{Skobeltsyn Institute of Nuclear Physics,\\Moscow State University,\\119992 Moscow, Russia}
\affiliation[c]{Institut f\"{u}r Mathematik und Institut f\"{u}r Physik,\\Humboldt-Universit\"{a}t zu Berlin,\\12489 Berlin, Germany}

\emailAdd{asmirnov80@gmail.com}
\emailAdd{smirnov@theory.sinp.msu.ru}

\abstract{New features of the {\tt Mathematica} code {\tt FIRE} are presented.
In particular, it can be applied together with the recently developed code {\tt LiteRed} by Lee
in order to provide an integration by parts reduction to master integrals for quite complicated
families of Feynman integrals. As as an example, we consider four-loop massless propagator integrals
for which {\tt LiteRed} provides reduction rules and {\tt FIRE} assists to apply these
rules. So, as a by-product one obtains a four-loop variant of the well-known three-loop computer 
code {\tt MINCER}.
We also describe various ways to find additional relations between master
integrals for several families of Feynman integrals.}

\newcommand{\be}{\begin{equation}}
\newcommand{\ee}{\end{equation}}
\newcommand{\bea}{\begin{eqnarray}}
\newcommand{\eea}{\end{eqnarray}}
\newcommand{\dd}{\mbox{d}}
\newcommand{\pa}{\partial}
\newcommand{\al}{\alpha}
\newcommand{\nn}{\nonumber}

\begin{document}

\maketitle
\flushbottom

\section{Introduction}

At the modern level of calculations in elementary particle physics, one often needs to 
evaluate thousands and millions of Feynman integrals. A classical approach is to apply the so-called
\textit{integration by parts (IBP) relations}~\cite{Chetyrkin:1981qh}  
(see Chapter~6 of \cite{Smirnov:2013ym} for a recent review)
and reduce all integrals to a smaller set,
the \textit{master integrals}\footnote{As it has been demonstrated in \cite{Smirnov:2010hn}, the number of master integrals is
always finite, so that, theoretically, this approach should be successful.}. 
A few years ago one of the present authors developed a program named
{\tt FIRE}~\cite{Smirnov:2008iw} performing reduction of Feynman integrals to
master integrals. Currently {\tt FIRE} written in {\tt Mathematica}
is one of a few public available codes (for other public products see
\cite{Anastasiou:2004vj,Studerus:2009ye,vonManteuffel:2012np,Lee:2012cn}) performing IBP reduction.

The purpose of this paper is to present {\tt FIRE4} -- the current {\tt Mathematica} version of 
{\tt FIRE}\footnote{All versions of {\tt FIRE} can be downloaded from \url{http://science.sander.su/FIRE.htm}.}.
Next section introduces the notation. In the following sections we are going to describe new features,
then we will analyze reasons on why {\tt FIRE} might work slowly and give some hints on resolving those issues.
In particular, we will explain how {\tt FIRE} can be applied together with the recently 
developed code {\tt LiteRed}~\cite{Lee:2012cn} by Lee
in order to provide an IBP reduction to master integrals for quite complicated
families of Feynman integrals. 
As as an example, we consider four-loop massless propagator integrals
for which {\tt LiteRed} provides reduction rules and {\tt FIRE} assists to apply these
rules. So, as a by-product one obtains a four-loop variant of the well-known three-loop computer 
code {\tt MINCER}~\cite{Gorishnii:1989gt}.
We also describe various ways to find additional relations between master
integrals for several families of Feynman integrals.

\section{Basic definitions}

Let us remind the notation we are going to use. Consider a Feynman integral as
functions of $n$ integer variables (indices),
\bea
  F(a_1,\ldots,a_n) &=&
  \int \cdots \int \frac{\dd^d k_1\ldots \dd^d k_h}
  {E_1^{a_1}\ldots E_n^{a_n}}\,,
  \label{eqbn-intr}
\eea
where the denominator factors $E_i$ are linear functions with respect to
scalar products of loop momenta $k_i$ and external momenta $p_i$, and dimensional regularization with
$d=4-2\epsilon$ is applied.

The integration by parts relations~\cite{Chetyrkin:1981qh} 
\bea \int\ldots\int \dd^d k_1 \dd^d k_2\ldots
\frac{\pa}{\pa k_i}\left( p_j \frac{1}{E_1^{a_1}\ldots E_n^{a_n}}
\right)   =0   \label{RR-intr}
\eea
can be rewritten in the following form:
\begin{equation}
\sum \al_i F(a_1+b_{i,1},\ldots,a_n+b_{i,n})
=0\,.
\label{IBP-intr}
\end{equation}
where $b_{i,j}\in \{-1, 0, 1\}$ and $\al_i$ are linear functions of $a_j$.

A classical approach is to separate all possible sets of indices into so-called sectors. Choosing a sector (one out of $2^n$) defines
for each index $a_i$ whether it is positive or non-positive. 
In fact, there are less than $2^n$ sectors -- indices corresponding to irreducible numerators are always non-positive.
A \textit{corner integral} in a sector is the one with indices equal to $0$ or $1$; each sector has a unique corner integral.

We say that a sector is \textit{lower} than another sector if all indices of integrals in the first one are smaller that corresponding
indices in the second one. Normally one tries to reduce Feynman integrals to those corresponding to lower sectors.
The reason for such a choice is that positive shifts always come 
with multiplication by the corresponding index, therefore relations written in sectors with negative values of indices do not 
depend on integrals with positive values of those indices. Moreover, integrals are simpler if more indices are non-positive.


The complexity of each integral corresponding to a given family (\ref{eqbn-intr}) is basically defined by 
the number of positive indices, and then the two non-negative numbers $N_+=\sum_{i\in \nu_{+}} (a_i-1)$ 
(the number of \textit{dots})
and
$N_-=-\sum_{i\in \nu_{-}} a_i$, where $\nu_{\pm}$ are sets of positive (negative) indices.

A sector is called \textit{trivial} if all integrals corresponding to sets of indices in this sector are equal to zero. The sector 
with all non-positive indices is always trivial. The conditions determining whether a sector is trivial are called
\textit{boundary conditions}.

A \textit{Laporta algorithm}~\cite{Laporta:2001dd} in a given sector is solving IBP's with a Gauss elimination after choosing an ordering.
The ordering choice and all details of the algorithm can be modified by the algorithm implementer.

\section{Combining {\tt FIRE} with {\tt LiteRed}}

A number of ideas and improvements to {\tt FIRE} were made due to Lee, both before and after his paper
on the code {\tt LiteRed} \cite{Lee:2012cn} appeared.

\subsection{Automatic determination of trivial sectors}
\label{adts}

The initial version of {\tt FIRE} used the {\tt RESTRICTIONS} variable in order to provide boundary conditions.
Now one can generate information on trivial sectors automatically. This is based on ideas presented in \cite{Lee:2008tj}.
IBP's form a tangent Lie algebra to the group of linear transformations of loop momenta with determinant one.
This yields such a statement that if applying IBP's to the corner integral result in it being equal to zero, then the sector is trivial.
This logic has been encoded in {\tt FIRE}, hence it can automatically detect boundary conditions in many 
cases.\footnote{This strategy of revealing trivial sectors was also suggested
by Pak \cite{Pak:2011xt}.}

Still, sometimes this does not detect all  the boundary conditions.  
For example, this is the case if we contract all the lines corresponding to non-positive indices (for a given sector)
and obtain a massless integral at the external momentum on the light cone. For details see section ``Scaleless integrals'' in \cite{Lee:2012cn}.
So, if one generates boundary condition automatically one might wish to mark missing trivial sectors manually.
This can be done by setting {\tt SBasisR[0,sector]} to {\tt True} after running the {\tt Prepare[]} command.
Here {\tt sector} is a list of {\tt 1} or {\tt -1} -- the signs of indices in the corresponding sector.

\subsection{Getting rid of a part of IBP's}

As it has been explained in \cite{Lee:2008tj}, the IBP's (before substituting indices) form a Lie algebra.
This knowledge lets one use less IBP's when performing reduction to master integrals.

This strategy has been implemented in the new version of {\tt FIRE}.
When working in a given sector, the IBP's are sorted so that the first ones maximally shift indices under the chosen ordering.
Then if a given integral is the highest one among integrals appearing in an IBP number $i$ (after substituting some indices),
then their is no need to apply IBP's with numbers $i$ or greater to this integral.

This approach decreases the number of redundant IBP's a lot and speeds up the reduction. The current version of {\tt FIRE} has this option by default set to {\tt True}.
The only reason to turn it off is if you are working with something different from classical IBP's that do not form a Lie algebra.

\subsection{Using LiteRed together with FIRE for IBP reduction}

The IBP reduction procedure present in {\tt LiteRed} \cite{Lee:2012cn}
is based on the approach of \cite{Lee:2008tj}
and aims to create reduction rules in all sectors. A set of reduction rules (or a basis in sector \cite{Smirnov:2006wh}) gives one a possibility to
reduce integrals in such a sector efficiently. No relations have to be solved anymore, for each integral (except for master integrals) one can quickly generate
a relation representing it in terms of lower integrals.

However even if bases exist in all sectors, the reduction speed depends a lot on the way those rules are applied. 
As it has been explained in \cite{Smirnov:2008iw}, there are basically two ways to perform reduction:

\begin{itemize}
\item Reduction variant 1:\\
the algorithm starts from lower sectors and lower integrals in them, step by step building the tables for the integrals with increasing complexity.
After each step each integral is represented in terms of master integrals.

\item Reduction variant 2:\\
the algorithm starts from higher sectors and higher integrals in them,
applies reduction rules to them and writes them into tables. On this pass nothing is substituted.
Each integral is represented in terms of few integrals that are ``a bit'' lower.
Then one makes a pass back, substituting in those tables from lowest to highest integrals.
During this pass each integral is represented in terms of master integrals.
\end{itemize}

Both approaches might work efficiently (in {\tt FIRE} we choose variant~2), however different approaches might
reduce performance a lot. The problem is that if one starts substituting table rules partially, the length of 
expressions starts to grow and is not limited any longer. And this also leads to a growth of coefficients' size and
problems with algebraic simplifications.

\begin{figure}
\includegraphics[width=1.\textwidth]{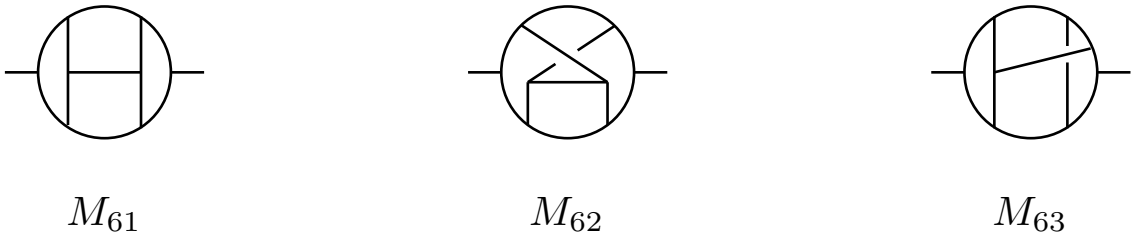}
\caption[]{Three most complicated four-loop propagator master integrals}
\label{baikov}
\end{figure}

Using {\tt FIRE} together with {\tt LiteRed} can make reduction much faster. To illustrate this feature 
let us choose an example of one of the most complicated families of massless four-loop propagator integrals.
This is the family $F(a_1,a_2,\ldots,a_{14})$ of integrals which includes the master integral $M_{62}$, in 
the notation of \cite{Baikov:2010hf} -- see Fig.~\ref{baikov}.

Using the choice of the loop momenta and numerators (associated with the first three indices)
as in \cite{Lee:Webpage} we have the following explicit expression for a general Feynman integral
of this family:
\bea
F(a_1,a_2,\ldots,a_{14})=
\int\ldots\int\dd^d l_1\ldots\dd^d l_4\frac{((l_4 + q)^2)^{-a_1}((l_2 -l_4)^2)^{-a_2}
}
{((l_1 - l_2 + l_3 - l_4 )^2)^{a_4}((l_1 - l_4 )^2)^{a_5}}
&& \nn \\ && \hspace*{-110mm}
\times 
\frac{((l_2 - l_3 + l_4)^2)^{-a_3}}{(( l_1 - l_2  )^2)^{a_6}((l_2 - l_3 )^2)^{a_7}((l_3 - l_4)^2)^{a_8}((l_1 + q )^2)^{a_9}((l_2 + q)^2)^{a_{10}}}
\nn \\ && \hspace*{-110mm}
\times 
\frac{1}{((l_3 + q)^2)^{a_{11}}(l_3^2)^{a_{12}}(l_4^2)^{a_{13}}(l_1^2)^{a_{14}}}\;.
\label{family62}
\eea

The {\tt LiteRed} rules for the family $F$ can be downloaded from 
\cite{Lee:Webpage}. (One takes the ``p4'' bases.)
To reduce a sample integral with these bases with {\tt LiteRed} one should launch {\tt Mathematica} and run:

$ $

\tt
SetDirectory[NotebookDirectory[]];

<$ $<LiteRed`;

SetDim[d];

Declare[\{l1, l2, l3, l4, q\}, Vector];

sp[q, q] = 1;

<$ $<"p4 dir/p4"

IBPReduce[j[p4, -2, 0, 0, 1, 1, 1, 1, 1, 1, 1, 1, 1, 1, 1], "file"]
\normalfont

$ $

To do the same with the use of {\tt FIRE} one needs to define propagators and other data first.

\tt
Internal = \{l1, l2, l3, l4\};

External = \{q\};

Propagators = \{(l4 + q)$^2$, (-l2 + l4)$^2$, 
    (l2 - l3 + l4)$^2$, 
    (l1 - l2 + l3\\ - l4)$^2$, (-l1 + l4)$^2$, 
    (l1 - l2)$^2$, (l2 - l3)$^2$, (l3 - l4)$^2$, 
    (l1 + q)$^2$, (l2\\ + q)$^2$, (l3 + q)$^2$, 
    l3$^2$, l4$^2$, l1$^2$\};

Replacements = \{q$^2$ -> 1\};

startinglist = 
  Flatten[Outer[(IBP[\#1, \#2] /. Replacements) \&,\\ Internal, 
    Join[Internal, \\External]], 1];
    
SYMMETRIES = \{\};

Get["FIRE\_4.0.0.m"];

Prepare[AutoDetectRestrictions -> True];
\cite{Lee:2012cn}
SaveStart["p4"];
\normalfont

$ $

Now on a clean kernel one can load the start file, the bases and evaluate the integral:

$ $

\tt
Get["FIRE\_4.0.0.m"];

LoadStart["p4", 1];

LoadLRules["p4 dir", 1];

Burn[];

EvaluateAndSave[\{\{1, \{-2, 0, 0, 1, 1, 1, 1, 1, 1, 1, 1, 1, 1, 1\}\}\}, \\
  "p4.tables"];
\normalfont

$ $

The tables can be loaded later on a clean kernel with

$ $

\tt
Get["FIRE\_4.0.0.m"];

LoadStart["p4", 1];

LoadLRules["p4 dir", 1];

Burn[];

LoadTables["p4.tables"];
\normalfont

$ $

\noindent so that the answer for the integral under consideration can immediately be retrieved within

$ $

\tt
F[1, \{-2, 0, 0, 1, 1, 1, 1, 1, 1, 1, 1, 1, 1, 1\}]
\normalfont

$ $

We ran some tests, comparing {\tt LiteRed} alone and {\tt FIRE} with the use of {\tt LiteRed} bases on the integrals
$F(-n, 0, 0, 1, 1,\ldots, 1)$
given by (\ref{family62})
for {\tt n = 1, 2, \ldots,} (with {\tt n = 0} it is a master-integral).
The results obtained are shown in Tab.~\ref{tc} (time is given in seconds\footnote{Time filled with ``?'' exceeds 18 days or one and a half million seconds.}, 3.07 GHz processor was used).

\begin{table}[!htb]
\centering
\tt
\begin{tabular}{|c|c|c|c|c|c|}
\hline
 n & 1 & 2 & 3 & 4 & 5
 \\
 \hline
 LiteRed & 943 & 9988 & 96318 & ? & ? 
 \\\hline
 FIRE {\normalfont with} LiteRed & 1251 & 2430 & 7044 & 27762 & 197823
 \\\hline
\end{tabular}
\normalfont
\caption{Sample evaluation time (in seconds).}
\label{tc}
\end{table}

\noindent As the tests show, a combined usage of these two programs is highly recommended.
We did not even try to run {\tt FIRE} on these complicated integrals alone -- it would take too much time.

Let us emphasize that the rules of reduction obtained within {\tt LiteRed} are similar in their character
to rules obtained by hand. This means that the code based on these rules is nothing but the four-loop
variant of the well-known package {\tt MINCER} \cite{Gorishnii:1989gt}. 
It is even more than that:  {\tt LiteRed} provides a reduction to true master integrals, while
the hand-made algorithm on which {\tt MINCER} is based reduces given integrals to master integrals {\em and}
some families of simple integrals which can be expressed explicitly in terms of gamma functions, so that
{\tt MINCER} provides a result for any three-loop massless propagator integral in an expansion in epsilon up to some
order. The rules obtained automatically with {\tt LiteRed} can be quite cumbersome. The above example shows that
their application within {\tt FIRE} turn out to be more effective that with {\tt LiteRed} itself.
(Well, at least within the current version of {\tt LiteRed}.)

Suppose now that we have succeeded to construct bases {\em almost} in all the sectors using {\tt LiteRed}.
Then it is even more important to turn to {\tt FIRE} and run it also to perform an IBP reduction in missing
sectors.

\section[Summary of smaller improvements in {\tt FIRE} since version 3.0]
{Summary of smaller improvements in {\tt FIRE} since version 3.0
\footnote{Version 3.0 was the first public version of {\tt FIRE} \cite{Smirnov:2008iw}.}}

\begin{itemize}
 \item {\tt FIRE 4} comes as a single file including {\tt IBP} and {\tt SBases} and required parts 
 of {\tt tsort} (see \cite{Pak:2011xt} and the discussion in section~\ref{tsort}). Since now it can be combined with {\tt LiteRed} 
 we removed the part related to the construction of Gr\"{o}bner bases from the code -- the {\tt LiteRed} bases are more efficient.

 \item Multiple speed and memory usage improvements.
 \item Multiple bug fixes.
 
 There is nothing much to tell about these two items, however we can state that in some cases the new {\tt FIRE} works a lot faster, and
many bugs have been fixed.
 
 \item The possibility to work with more than fourteen indices. (This is the number of indices for families of four-loop propagator integrals.)
 
 This might be complicated for the {\tt Mathematica} version, but at least it works in principle.
 
 \item {\tt DatabaseUsage 4} and {\tt MemoryLimit}.
 
 {\tt FIRE} uses database engines to store tables on a hard disk. Depending on its settings it might store more or less 
 (the more one stores, the less RAM one needs, but the more performance degrades).
 The first public version of {\tt FIRE} had the {\tt DatabaseUsage} setting that could be changed from  {\tt 0} (no usage) to {\tt 3} (maximum).
 Currently there is also possible value {\tt 4}, but what is more important, there is a setting {\tt MemoryLimit}. 
 If it is set, {\tt FIRE} automatically increases {\tt DatabaseUsage} upon reaching the limit (measured in megabytes).
 
 \item The possibility to choose master integrals.
 
  Sometimes one wants to choose master specific integrals. For example there are two master integrals in a sector.
  One of them is normally the corner integral in the sector, but the second integral choice might be different.
  If no priority is set, {\tt FIRE} chooses the second master integral itself. However, there might be some reasons 
  to change this choice. For example, one of those integrals can be calculated easier than the other.
  
  This priority can be defined by the {\tt MakeMaster} with two options -- the integral and the priority (a positive integer). 
  The integrals marked with {\tt MakeMaster} are preferred to the ones not marked and are compared between each other by the priority set.
 
 \item The possibility to construct reduction bases automatically.
 
  The first public version of {\tt FIRE} came together with the possibility to build reduction bases in sectors 
  (based on the Gr\"{o}bner bases approach). Currently we can claim this approach (within {\tt FIRE}) to be inefficient: 
  the rules constructed by the
  {\tt LiteRed} program work much better. However, in many sectors one can now construct reduction rules only with the use of {\tt FIRE}, 
  and they can work better that the ones from {\tt LiteRed}.
  
  The idea is that if there are no master integrals in a given sector, there is a chance of finding a single IBP that
  can throw all integrals from this sector. Of course, one does not fix an ordering in this case; the algorithm analyses all IBP's
  and tries to find an ordering such that such a reduction will be possible.
  A search for these bases can be initiated by the {\tt BuildAll[region]} command.
 
\end{itemize}

To demonstrate bases construction let us consider a simple massless box diagram.

$ $

\tt
Get["FIRE\_4.0.0.m"];

Internal = \{k\};

External = \{p1, p2, p4\};

Propagators = \{-k$^2$, -(k + p1)$^2$, -(k + p1 + p2)$^2$, -(k + p1 + p2 + 
       p4)$^2$\};
       
Replacements = \{p1$^2$ -> 0, p2$^2$ -> 0, p4$^2$ -> 0, p1 p2 -> -S/2, 
   \\p2 p4 -> -T/2, p1 p4 -> (S + T)/2, S -> 1, T -> 1\};
       
PrepareIBP[];

startinglist = 
  Flatten[Outer[(IBP[\#1, \#2] //. Replacements) \&, \\Internal, 
    Join[Internal, External]], 1];
    
\normalfont

$ $

Now we do not provide diagram symmetries, but simply autodetect boundary conditions, 
construct bases automatically, save the file and quit the kernel.

$ $

\tt
Prepare[AutoDetectRestrictions -> True];

BuildAll[\{0, 0, 0, 0\}];

SaveSBases[``box''];

Quit[];

\normalfont

$ $

The input of {\tt BuildAll} specifies that all indices can be both positive, or non-positive. If one puts {\tt -1} instead of {\tt 0}
in some place, then the corresponding index can be only negative.

One can notice that the code has built bases in $3$ sectors out of $11$. That is not much, but sometimes it works better, 
and if one is going for maximum performance, one should not forget about those things.

Now on a clean kernel the bases can be loaded and we can go for an evaluation:

$ $

\tt

Get["FIRE\_4.0.0.m"];

LoadSBases[``box'',2];

Burn[];

F[2,\{2,2,2,2\}];
\normalfont

$ $

One can quickly see that the result depends on three master-integrals:

$ $

\tt
\{G[2, \{1, 1, 1, 1\}], G[2, \{0, 1, 0, 1\}], G[2, \{1, 0, 1, 0\}]\}
\normalfont

$ $

The second and the third integral are identical, but {\tt FIRE} had no information to retrieve that.
To give this information to {\tt FIRE} one could provide global symmetries (do not forget to do it in practice), however sometimes
identical integrals cannot be located by global symmetries of the diagram.

\section{Getting rid of extra master integrals}
\label{gremi}

FIRE cannot directly identify master integrals unless they are equivalent to each other under a global symmetry of the diagram.
Hence after running a sample reduction one might find that the tables contain too many master integrals. (Their list can be produced
by the {\tt GetII/@IrreducibleIntegrals[]} command). 

If we deal with several families of Feynman integrals (for example, relevant
to a given physical problem) we take the union of the sets of the master integrals corresponding
to the individual families. (It often happens that some master integrals belong to different
families.) Then we could try to find relations in this united family. 
In this section, three tools of finding such relations are presented.

In fact, in certain situations, it is clear that extra relations {\em should exist}.
For example, one can observe that some denominators in relations expressing integrals in terms of
master integrals turn out to be quadratic in $d$, or even unfactorizable polynomials of a higher degree.
However, the convergence analysis of general Feynman integrals shows that the poles in $\varepsilon$ belong to the real axis,
so that such complicated factors in the denominator should not be present.
Indeed, after finding extra relations between the master integrals of the current family,
these spurious denominators disappear.
 
\subsection{{\tt tsort}}
\label{tsort}

The simplest way to find extra relations is to use 
an integral identification algorithm by Pak~\cite{Pak:2011xt} based on
alpha-representation. We will refer to this algorithm as {\tt tsort}.
This code creates alpha-representations of integrals in a given list, puts them into certain canonical forms, then
finds equivalents between integrals. The relations can be produced in a form of {\tt Mathematica} rules by their
{\tt FindRules} command or directly saved into a file with the {\tt WriteRules} command. Such a file with rules can be used in
consequent reductions.

If we return to the box example above, we can now run (the propagators, momenta and replacements have to be provided):

$ $

\tt
list=GetII /\@ IrreducibleIntegrals[];

Internal = \{k\};

External = \{p1, p2, p4\};

Propagators = \{-k$^2$, -(k + p1)$^2$, -(k + p1 + p2)$^2$, -(k + p1 + p2 + 
       p4)$^2$\};
       
Replacements = \{p1$^2$ -> 0, p2$^2$ -> 0, p4$^2$ -> 0, p1 p2 -> -S/2, \\
   p2 p4 -> -T/2, p1 p4 -> (S + T)/2, S -> 1, T -> 1\};

WriteRules[list, "box"];
\normalfont

$ $

Now if we run the task again on a clean kernel with

$ $

\tt

Get["FIRE\_4.0.0.m"];

LoadSBases[``box'',2];

Burn[];

LoadRules[``box'',2];

F[2,\{2,2,2,2\}];
\normalfont

$ $

\noindent then we will obtain a result with two master-integrals. Or if we already have a result, we can run

$ $

\tt

list=GetII /\@ IrreducibleIntegrals[];

res=res/.FindRules[list];

\normalfont

$ $

As a second example, let us briefly characterize the situation with the master integrals corresponding to
the three families associated with the master integrals $M_{61},M_{62},M_{63}$ \cite{Baikov:2010hf} -- see Fig.~\ref{baikov}.
When we work with the families individually, using either {\tt LiteRed} or {\tt FIRE}, we encounter
18, 16 and 21 master integrals, correspondingly. Let us, for definiteness, choose the family associated with
$M_{61}$ as the main family. Then  {\tt tsort} shows that 12 of 16 master integrals of the $M_{62}$ family
and 16 of 21 master integrals of the $M_{63}$ family
can be mapped to the $M_{61}$ family. Moreover, 2 master integrals of the $M_{63}$ family
can be mapped to the $M_{62}$ family. Altogether, there are 25 master integrals in the union of these three families.
The three missing master integrals in the whole family of four-loop massless propagator integrals
are simpler factorisable integrals -- see \cite{Baikov:2010hf}.

As one can see, in order to find extra relations between supposed master integrals, one has to have a list of them. In order to create such a list,
one has to perform a reduction. In order to make this reduction fast enough one needs the knowledge of equivalence relations between 
master integrals. To get out from this loop (if everything is too slow) we suggest first running sample reduction jobs with 
integrals simpler than the ones required for the physical calculation. After that one obtains a subset of master integrals,
finds equivalents between them and then runs jobs that are more complicated. This procedure can be repeated iteratively.

One problem is that this canonical form approach cannot work properly with integrals with irreducible numerators (negative indices).
In order to have a list that has only integrals with dots, one can use the {\tt MakeMaster} command as explained earlier,
marking all integrals with one dot (or even two or three dots) as preferred for being masters in problematic sectors.

Let us emphasize that this code {\tt tsort} can effectively be applied even within one family of Feynman integrals,
while the next two tools are more important when dealing with two and more families.

\subsection{Using symmetries to find extra relations}

\begin{figure}
\includegraphics[width=.8\textwidth]{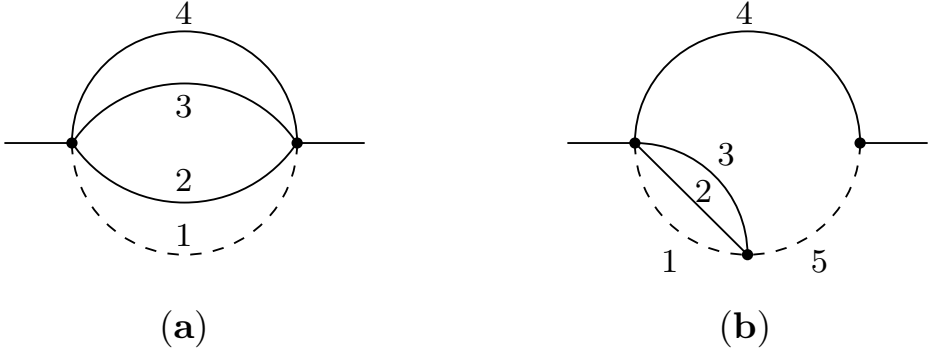}
\caption[]{The integral $I_{11}$ ({\bf a})  and the auxiliary diagram ({\bf b}) used for its reduction.
Solid (dotted) lines denote massive (massless) propagators, the external momentum is
at $p^2=m^2$.
}
\label{redm2}
\end{figure}

Let us consider the diagram  of  Fig.~\ref{redm2}a with the external momentum at $p^2=m^2$.
This is an integral with the numerator $k\cdot p$ where $k$ is the momentum of the massless line.
It is denoted by $I_{11}$ in \cite{Laporta:1996mq}
and belongs to the set of the master integrals contributing to the three-loop $g-2$ factor.
It was present, in addition to the corresponding 
master integral $I_{10}$ without numerator. Indeed, if one runs an IBP reduction for such integral with numerators
one obtains two master integrals in the upper sector, i.e. with positive four indices associated with the
propagators.
Later it was observed in \cite{Laporta:1997zy} that $I_{11}$ is a linear combination of the integrals $I_{14}$ and $I_{18}$
(or, $J_{14}$ and $J_{18}$ in the notation of \cite{Laporta:1997zy}).
This linear connection was present in  \cite{Laporta:1997zy} with the coefficient at $I_{14}$ expanded in $\varepsilon$ up to a certain power,
rather than exactly at general dimension $d$.
In \cite{Lee:2010ik} this relation was presented at general $d$:
\be
I_{11}=\frac{2d-5}{2(d-2)}I_{14}-\frac{1}{4}I_{18}\;.
\label{I11red}
\ee
In the notation of \cite{Lee:2010ik}, we have $I_{14}=G_{4,4}$ and $I_{18}=G_{3}$. 
 
The relation (\ref{I11red}) can be derived using a symmetry of Feynman integrals.
In the case of Feynman integrals connected with  $I_{11}$ no symmetry can help
to reduce the number of the master integrals and we have two master integrals in the highest sector.
However, if our goal is to reduce the number of master integrals for several families
of Feynman integrals considered together we can profit from a symmetry.
To reduce $I_{11}$ it is enough to consider the family of Feynman integrals corresponding
to the graph Fig.~\ref{redm2}b. 
In particular, we have $F(1,2,1,1,1)-F(1,1,2,1,1)=0$.
However this relation is automatically satisfied after applying an IBP reduction.
It turns out that the missing relation can be revealed at the next level of indices:
if we reduce $F(1,2,1,2,1)-F(1,1,2,2,1)=0$ to the master integrals we indeed obtain
an equation which leads to (\ref{I11red}).

\subsection{Using differentiation to find extra relations}

Let us turn to one more way to obtain extra relations between master integrals 
\begin{figure}
\includegraphics[width=.8\textwidth]{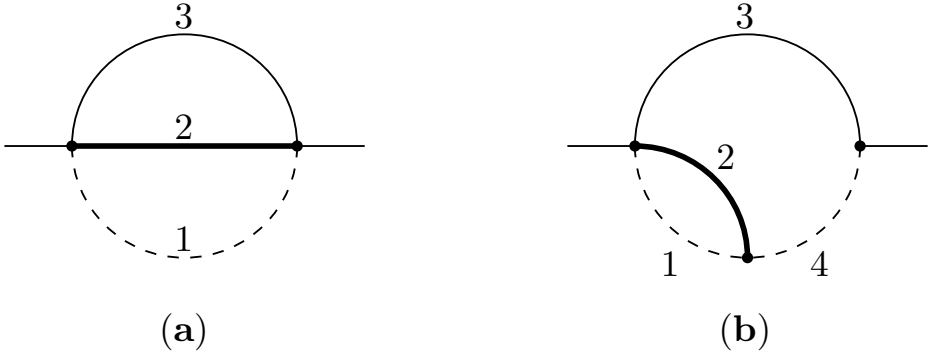}
\caption[]{A two-loop diagram with the masses $0,M,m$ ({\bf a}) and the auxiliary diagram ({\bf b}) used for its reduction.
Dotted, thick and thin lines denote propagators with the masses $0,M$ and $m$, correspondingly. The external momentum is
at $p^2=m^2$.
}
\label{redm}
\end{figure}
using the diagram of Fig.\ref{redm}a considered at $p^2=m^2$ as an example.
 
In \cite{Kalmykov:2011yy} it was shown that a linear combination of the three master integral in the highest sector
of the family of integrals associated with Fig.\ref{redm}a  
is a function which is expressed in terms of gamma functions at general $d$. It was observed that this function
 is given by a two-loop vacuum diagram with two zero masses.
 This analysis was based on explicit representations of Feynman integrals for Fig.\ref{redm}a 
 in terms of hypergeometric functions and recurrence relation between hypergeometric functions.
Later such relation was derived in \cite{Kniehl:2012hn} using a trick with an introduction of an auxiliary mass
and, in \cite{Kalmykov:2012rr} using recurrence relations between MB integrals representing Fig.\ref{redm}a.
 
It turns out that this extra relation can be derived using IBP reduction and differentiation with respect to $M$.
Similarly to the previous example, let us consider a diagram with one more propagator depicted in
Fig.\ref{redm}b. Using the numbering in this figure let us consider $F(1, 1, 1, 2)$. First, we reduce it to master integrals.
(In order to solve IBP relations, we need to introduce a fifth index connected with an extra irredcible
numerator. However, it does not appear in the relations that follow.)
Second, we use the fact that $F(1, 2, 1, 2)=-\frac{\pa}{\pa M^2}F(1, 1, 1, 2)$. 
So, we reduce $F(1, 2, 1, 2)$ to master integrals and equate the difference between the
corresponding result and the derivative of the result of reduction of $F(1, 1, 1, 2)$. 
We straightforwardly arrive at the relation
\bea
(3 d-8) F(1, 1, 1, 0) +  4 m^2 F(1, 1, 2, 0) 
&& \nn \\ && \hspace*{-50mm}   
 + 2 M^2 F(1, 2, 1, 0)+(2 - d) F(1, 1, 0, 1)=0
\eea
which is noting but the additional relation of \cite{Kalmykov:2011yy}.

\section{{\tt FIRE} works slowly. Why?}

There can be multiple reasons.

\subsection{Proper input}

Be sure to provide proper input for {\tt FIRE}. To do that, one has to keep in mind that one has to specify
\begin{itemize}

\item A complete set of IBP's. If one provides less IBP's than expected, the code will result in multiple extra master integrals and work very slowly. 
 \item Boundary conditions. Either specify them with the {\tt RESTRICTIONS} setting, or run {\tt Prepare} with {\tt AutoDetectRestrictions->True}. 
 In the second case one might skip 
 some boundary conditions, specify the remaining trivial sectors manually if needed (see Section~\ref{adts}).
 
 If there are trivial sectors, that are not marked as trivial, {\tt FIRE} will finally detect that they are trivial, 
 however it takes really much time.
 
 \item Global symmetries. Specifying proper global symmetries might increase the reduction speed a lot. However, please, keep in mind that
 one should specify not only generators of the symmetry group, but all non-trivial permutations.
  
 \item A proper syntax to evaluate multiple Feynman integrals of a given family is the {\tt EvaluateAndSave} command. Do not try to calculate them one by one with the {\tt F[\ldots]} syntax.
 
\end{itemize}

\subsection{Reduction bases}

\begin{itemize}

 \item Building reduction bases automatically might speed up things a lot. Do not forget about the {\tt BuildAll} command.
 The result also depends on the way the IBP's are chosen. It is always correct, but the number of sectors where the bases are constructed can differ.
 A ``rule of the thumb'' to produce best IBP's is the following: choose (partially intersecting) loops on the Feynman diagram and corresponding loop momenta.
 Now when one is differentiating by some loop momenta, try to multiply only by momenta going through lines in the corresponding loop.

 \item Using the new {\tt LiteRed} package together with {\tt FIRE} speeds up reduction a lot. The {\tt LiteRed} package
 has to be downloaded separately and used to create reduction bases that can be used from within {\tt FIRE}.

\end{itemize}

\subsection{Computer issues}

The {\tt Mathematica} version of {\tt FIRE} cannot work in parallel mode, hence the number of cores does not influence reduction speed.
The amount of RAM is much more important for efficient reduction. One has no analyze the job and monitor whether it runs out of RAM.
If so, one has to specify a higher {\tt DatabaseUsage} setting, set a {\tt MemoryLimit} or think of moving to another computer.

If working already with a nonzero {\tt DatabaseUsage} one has to provide a proper path for storing the database.
If one is working on a cluster machine, he/she should definitely set this path to be a local hard disk and not a disk somewhere on the network.
The hard disk should also be big enough and fast enough and preferably not used by other processes. An SSD (solid state drive) is also a good choice.

While one uses the database, the operating system usually caches a part of it in RAM.
This needs less RAM than with direct RAM usage and after the RAM is completely filled with cache, 
the program is not going to become too slow (as it happens with swapping). However,
a certain performance degradation happens after this point.

One should also keep in mind that long calculations should better not be launched within the {\tt Mathematica} frontend. 
A batch job is recommended. Or if one is using a personal computer, he can at least launch the {\tt Mathematica} kernel with the {\tt math} command
and run the job from out of there.

\subsection{Extra master integrals}

If the reduction goes fast enough, but it takes too much time to perform the final substitutions, then this is the problem
with extra master integrals. They have to be somehow related with each other to make the reduction faster.
The details have been explained in Section~\ref{gremi}.

\section{Conclusion}

In this paper we presented a new version of {\tt FIRE}, a program used to perform
reduction of Feynman integrals to master integrals. 
This version has some tools such as {\tt tsort} \cite{Pak:2011xt} by Pak already inside, 
but we also recommend to use it in conjunction with the {\tt LiteRed} \cite{Lee:2012cn} package by Lee.
This paper can be also considered as a user guide on how to use them together in an efficient manner.

In future we are planning to release a {\tt c++} version of {\tt FIRE} that is currently in development stage. 
All the approaches listed in this paper can also be used in {\tt c++} {\tt FIRE} so that everything is going to be compatible.

\section*{Acknowledgements}
 
This work was supported by the Russian Foundation for Basic Research through grant 11-02-01196.
The work of V.S. was also supported by the Alexander von Humboldt
Foundation. We are greateful to Pavel Baikov, Konstantin Chetyrkin, Roman Lee, Alexey Pak and Matthias Steinhauser for helpful discussions and careful
reading of draft versions of this paper.

\bibliographystyle{JHEP}
\bibliography{FIRE-new,asmirnov}
\end{document}